# Machine Learning Sensor Fusion for Water Vapor Radiometry based Tropospheric Delay Calibration

Tirupati K. Sridharan(1), Yoshiharu Asaki(2) & Brian Svoboda(1)
(1) National Radio Astronomy Observatory
(2) National Astronomical Observatory of Japan

***Abstract*—The next-generation Very Large Array (ngVLA) intends to use 22 GHz water vapor radiometry (WVR) to correct for tropospheric phase fluctuations in its high-frequency bands. However, current delay-retrieval methods do not reach the required accuracy and reliability. As a remedy, we propose a machine-learning based retrieval concept and a preliminary architecture, framed as heterogeneous, multi-rate sensor fusion. Atmospheric reanalysis profiles and surface weather set the context in which per-antenna WVR spectra are used to estimate excess path by a Kalman filter whose parameters are learned from high-cadence calibrator phases. The method is being developed, with planned first validation on the 183 GHz WVR system of the Atacama Large Millimeter/submillimeter Array (ALMA).**

## I. Introduction

The ngVLA requires the per-antenna tropospheric excess path to be estimated to ~0.1 mm on seconds time scales, to attain 28-32 dB image dynamic range in bands 5 & 6 (30.5–116 GHz) [1]. While 22 GHz water vapor radiometry has been adopted to achieve this with lower calibration overheads than fast-switched gain calibration, current WVR-based excess-path correction methods fall short of the requirements and approaches beyond the traditional path retrieval paradigm may be needed [2].

Traditional retrievals interpret the sensed 22 or 183 GHz water line at each antenna through a simple atmospheric model; e.g., the ALMA 183 GHz phase-correction system fits a thin single-layer model to its four WVR channels [3], ignoring the vertical structure of the atmosphere and surface weather; residual errors are reduced empirically by scaling the solutions against calibrator phases [4]. Inverting a few-channel spectrum for excess path, with unknown vertical profiles of temperature and water vapor, is not a well-posed problem in general, which may partly explain the variable success of 22 GHz systems.

We will use the ERA5/MERRA-2 general circulation model (GCM) based reanalyses [5][6] to provide the missing vertical profile, and weather sensors for surface information, and to learn the retrieval from data. Archival ALMA data offer a proof-of-concept path, with future 22 GHz VLA/ngVLA extensions.

## II. Problem Framing and Data

The problem is framed as multi-rate, heterogeneous sensor fusion, closely related to data assimilation: sensors of different accuracy, cadence, and spatial scale are combined, through radiative transfer physics, to estimate a single quantity, the excess path length. Table I lists the inputs for the ALMA proof-of-concept phase, which we focus on here as the first step. The *am* radiative transfer model [7] is used to compute the coarse model spectrum and its Jacobian with respect to PWV, using the reanalysis profiles and the surface weather data as inputs.

Table I. Inputs for the ALMA Proof of Concept.

| Input | Cadence, Resolution | Role |
|---|---|---|
| N-channel WVR spectra (183 GHz, N=4) | 1.2 s, per antenna | Fine-cadence observation |
| Surface weather (temperature, pressure, humidity) | Seconds, per antenna | Updates profile near the ground; input to learned correction |
| Reanalysis profiles of temperature and water vapor (ERA5/MERRA2) | 1–3 h, 30–60 km | Coarse background atmospheric state |
| Calibrator phase (bandpass scans) | 6 s in training; about 3 min in operation | Ground truth; anchors |

The ground truth is the antenna-based phase measured on bright bandpass calibrators, available every 6 s within ~3 min-long scans. With 40 antennas and 100 such scans per year over 15 years, the ALMA archive provides ~$10^3$ array data segments containing ~$10^5$ antenna data segments for training. In operation only sparse calibrator measurements ("anchors"), ~3 min apart, are available. The task is to reconstruct the excess path between anchors using the coarse GCM data, and fine cadence WVR and weather data. The current scope is offline correction.

## III. State-Space Formulation

A kinematic Kalman filter approach is proposed for estimating the latent state $x_t$, defined as the excess path of one antenna relative to the mean of the antennas sharing a GCM reanalysis grid cell, which for ALMA is the whole array. All quantities are cast into differenced forms, because interferometric phase calibration is differential.

### A. Transition Model

The augmented latent state $X_t = [x_t, \dot{x}_t]$, the path and its rate of change, evolves as

$$X_t = FX_{t-1} + w_t , \qquad (1)$$

where $F$ is the transition matrix, with a constant-velocity model as a simple starting placeholder, and $w_t$ is process noise with covariance $Q$, set by a single intensity $q$. A future Kolmogorov turbulence-based transition model, with temporal structure function exponents of 5/3 and 2/3 [8], will capture more physics.

## B. Observation Model

The observation is the measured WVR spectrum of the antenna minus the group-mean spectrum, $S_t$ (differential). As the GCM-based model spectrum is common to the group for ALMA it cancels in the corresponding model difference spectrum, and the reanalysis information enters only through the Jacobian:

$$S_t \approx J_t\, x_t + g(u_t;\, \theta) + v_t. \tag{2}$$

Here $J_t$ is the *am*-computed sensitivity of the spectrum to excess path, for a hybrid profile (GCM reanalysis and surface weather $u_t$); $g$ is a learned weather correction with parameters $\theta$; and $v_t$ is observation noise with covariance $R$. At each time step the filter predicts the state as $\hat{X}_{t|t-1} = F\, \hat{X}_{t-1|t-1}$, the expected value under (1), and corrects the prediction using the new spectrum,

$$\hat{X}_{t|t} = \hat{X}_{t|t-1} + K_t\, [S_t - J_t\, \hat{x}_{t|t-1} - g(u_t;\, \theta)], \tag{3}$$

where $S_t$, $J_t$, and $g$ are four-element vectors (one entry per channel), $\hat{x}_{t|t-1}$ is the predicted path (the first component of $\hat{X}_{t|t-1}$), and the 2×4 Kalman gain $K_t$, chosen to minimize the estimation error, weights the prediction against the measurement according to $Q$ and $R$, which are learned. Antennas on the longer ngVLA baselines not sharing grid cells are referenced to a single antenna, and $S_t$ is formed by differencing their WVR − GCM difference spectra. This case is not dealt with further here.

With the GCM model spectrum cancelled, the filter incorporates the absolute atmospheric state only through $J_t$. The correction $g$, a learned function of surface weather $u_t$, partly restores the absolute dependence. The ALMA WVR scaling factors of [4] correlate with PWV and, more weakly, with pressure and temperature, which motivates incorporating the actual conditions through $g$.

## C. Role of the Reanalysis Profiles

Different vertical profiles can give nearly the same few-channel WVR spectrum. The reanalysis data supply the profile about which the spectrum is linearized. This allows inclusion of the vertical distribution of water vapor and temperature to first order through $J_t$, otherwise only poorly captured by the few-channel WVR spectrum. The larger number of channels in future ngVLA WVRs can capture the vertical structure better.

## D. Detrending Between Anchors

Within each antenna data segment, the calibrator-derived target excess path $y_t$, the spectra, and the filter output each have the linear trend between the 30 s anchor averages at the segment ends removed. This removes the dominant systematic error in the calibrator phases, a slow instrumental drift, and a linear trend of the atmosphere, which the anchors already supply. The delivered path is the detrended filter output plus the linear interpolation between the anchors, matching them exactly.

# IV. Learning and Validation

The noise parameters $q$ and $R$, and $\theta$ for weather correction $g$, are learned by minimizing, over the training segments,

$$L = L_{\text{sup}} + \lambda_{\text{p}}\, L_{\text{phys}} + \lambda_{\text{c}}\, L_{\text{cont}}. \tag{4}$$

The supervised term $L_{\text{sup}}$ is the summed squared difference between the estimate $\hat{x}_t$ ($\equiv \hat{x}_{t|t}$) and the calibrator-derived target $y_t$. The other two terms are regularizers that encode prior physical knowledge: $L_{\text{phys}}$ penalizes the residual of (2), $S_t - J_t\, \hat{x}_t - g(u_t;\, \theta)$, and $L_{\text{cont}}$ penalizes step-to-step changes in $\hat{x}_t$, with $\lambda_{\text{p}}$ and $\lambda_{\text{c}}$ being hyperparameters. Gradients are obtained by automatic differentiation through the unrolled filter (backpropagation through time). The Jacobians provided by the non-differentiable *am* model enter as fixed coefficients.

The data volume supports ~$10^2$–$10^3$ learned parameters, so the number of parameters in $\theta$ must be small and the main risk may be overfitting. Accuracy will be assessed by reconstruction from the end anchors alone on held-out array segments, scored as the rms of the (6 s target – estimate) difference, under $k$-fold cross-validation ($k \geq 5$) split by time. As all antennas of an array segment share one coarse scale atmosphere, they are kept in the same fold. An ablation study discarding $g$ will show whether a learned weather correction improves on a linear Kalman filter with learned noise parameters. Training is expected to be computationally light, with the dominant cost being the *am* evaluation at fine cadence, once for each time step, offline.

# V. Outlook

This is a first presentation of the concepts and a preliminary architecture for a machine learning approach to tropospheric delay calibration using WVRs, to be refined for implementation. Open questions include reconstruction of the absolute path, the reference-antenna treatment for ungrouped ngVLA antennas, and the turbulence-based transition model. An ALMA proof-of-concept phase paves the way to the ngVLA extension and could bring operational benefit to ALMA. A validated method could broadly apply to other current and future interferometers equipped with WVRs and surface weather sensors.

# Acknowledgment

Helpful exchanges with S. Bhatnagar, K. Massingill, and B. Mason are acknowledged. This work partly used AI (Claude) for learning, research, and discussions, and for summarizing the discussions for review. The National Radio Astronomy Observatory is a facility of the U.S. National Science Foundation operated under cooperative agreement by Associated Universities, Inc.